\documentclass{article}
\usepackage{spconf,amsmath,graphicx}
\usepackage{amsmath,amssymb,amsfonts}
\usepackage{cite}
\usepackage{url}
\DeclareMathOperator*{\argmax}{argmax}
\usepackage[usenames,dvipsnames]{color}

\title{Deep Joint Source-Channel Coding for Wireless Image Transmission with Adaptive Rate Control}
\name{Mingyu Yang, Hun-Seok Kim
\thanks{This work was funded in part by DARPA YFA \#D18AP00076 and NSF CAREER \#1942806.}}
\address{University of Michigan, Ann Arbor, MI, USA}
\begin{document}
%
\newcommand{\figstructure}{
  \begin{figure*}
    \centering
    \includegraphics[width=0.9\linewidth]{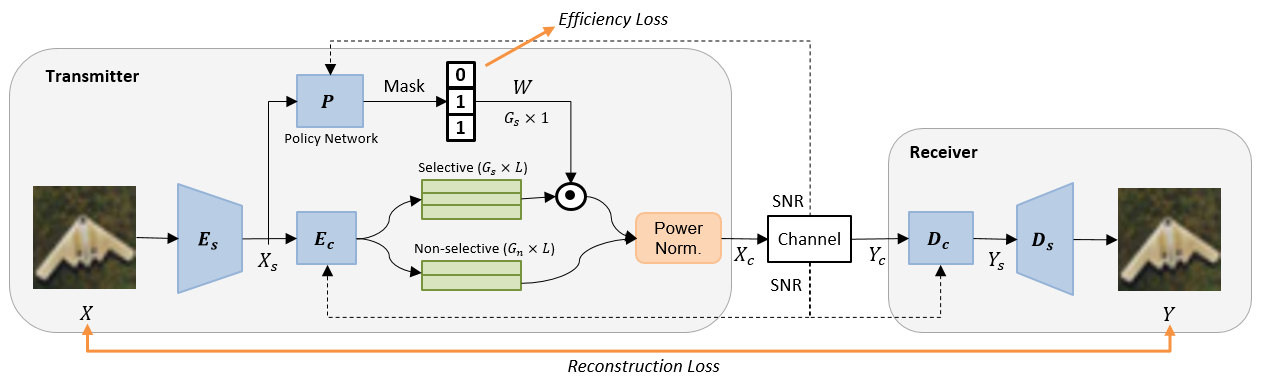}
    \caption{Overall structure of the proposed deep JSCC with rate control. Multiple rates are achieved by activating different subsets of the selective features. The mask is a thermometer code and is regularized by an efficiency loss.}
    \label{fig:fig_structure}
  \end{figure*}
}

\newcommand{\figadap}{
  \begin{figure}[!t]
    \centering
    \includegraphics[width=0.87\columnwidth]{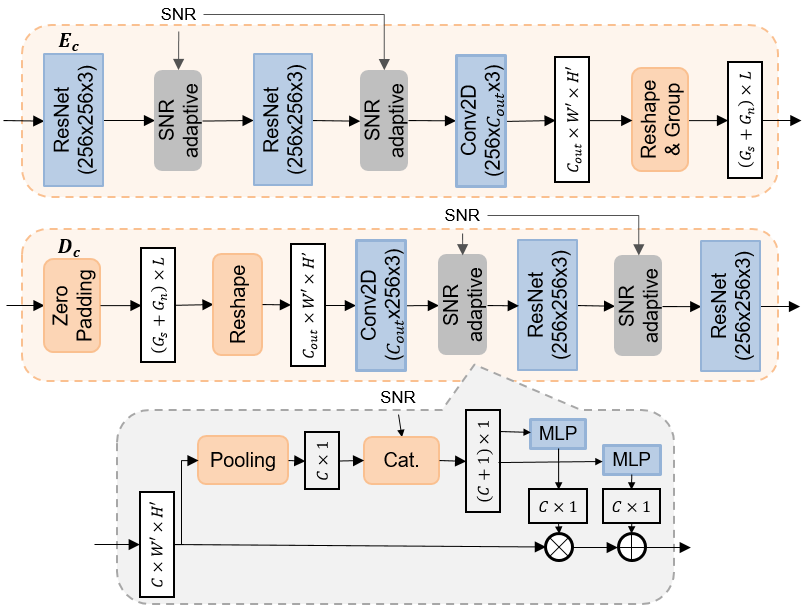}
    \caption{The structure of the channel encoder (top), channel decoder (middle) and SNR adaptive module (bottom). The parameters of DNN layers are in the format of \textit{input channel} $\times$ \textit{output channel} $\times$ \textit{kernel size}. All layers have a stride of 1. }
    \label{fig:fig_adaptive}
  \end{figure}
}

\newcommand{\figpolicy}{
  \begin{figure}
    \centering
    \includegraphics[width=0.9\columnwidth]{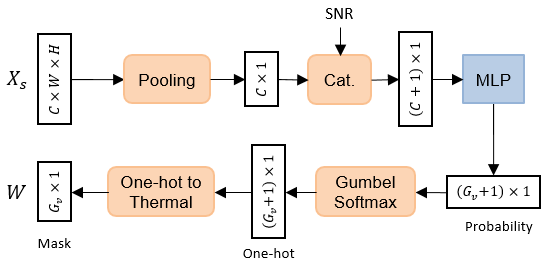}
    \caption{Structure of the policy network}
    \label{fig:fig_policy}
  \end{figure}
}

\newcommand{\figResPSNR}{
  \begin{figure}[t]
    \begin{minipage}[b]{0.48\linewidth}
       \centering
       \centerline{\includegraphics[width=4.8cm]{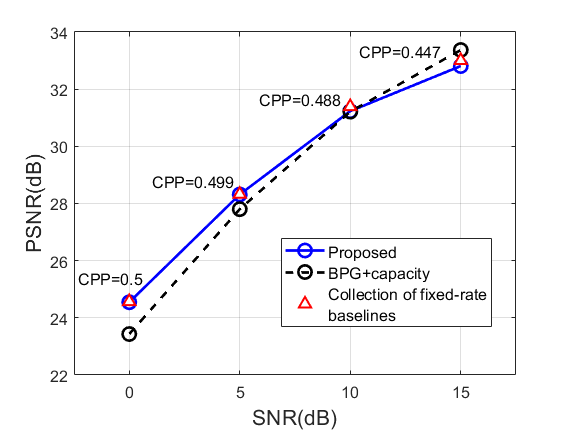}}
       \centerline{(a) $\alpha=5\times 10^{-4}$}\medskip
    \end{minipage}
    \hfill
    \begin{minipage}[b]{.48\linewidth}
       \centering
       \centerline{\includegraphics[width=4.8cm]{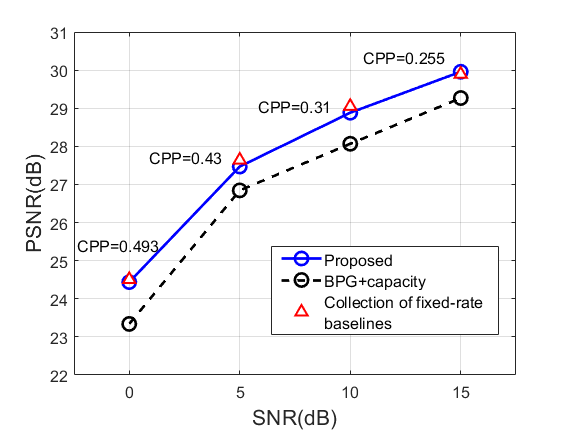}}
       \centerline{(b) $\alpha=1.5\times 10^{-3}$}\medskip
    \end{minipage}
\caption{Attainable PSNR comparison between the proposed method and baselines.}
\label{fig:fig_results_psnr}
  \end{figure}
}
\newcommand{\figResRate}{
  \begin{figure}[!t]
    \centering
    \includegraphics[width=0.8\columnwidth]{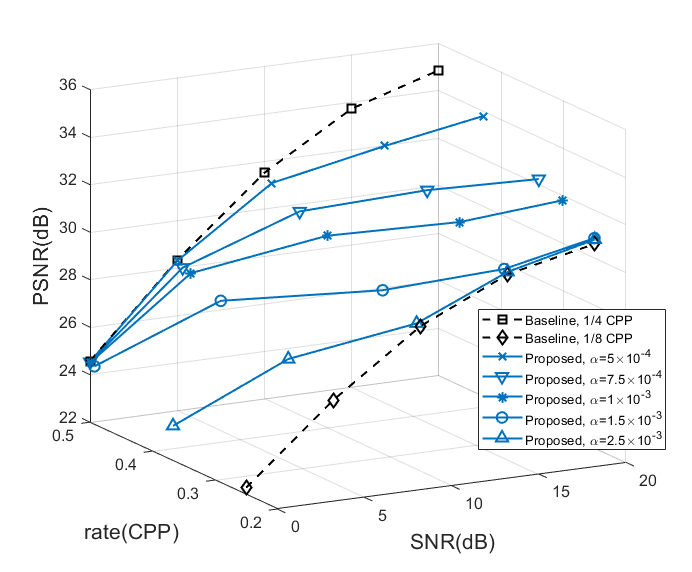}
    \caption{The average rate and average PSNR of the proposed method for different SNRs and $\alpha$'s.}
    \label{fig:fig_results_rate}
  \end{figure}
}
\newcommand{\figResVar}{
  \begin{figure}[!t]
    \centering
    \includegraphics[width=0.8\columnwidth]{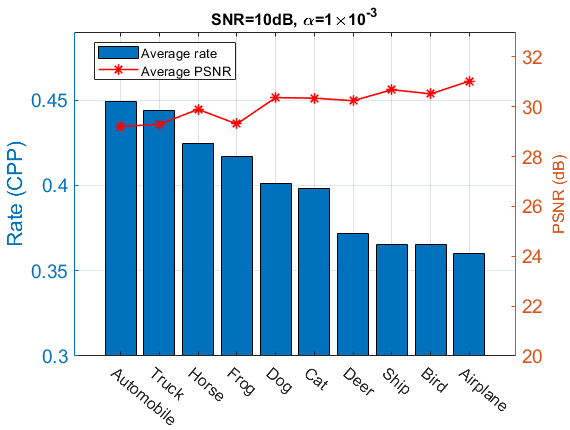}
    \caption{The average rate for different classes in CIFAR-10. }
    \label{fig:fig_results_variance}
  \end{figure}
}

\maketitle
\begin{abstract}
We present a novel adaptive deep joint source-channel coding (JSCC) scheme for wireless image transmission. The proposed scheme supports multiple rates using a single deep neural network (DNN) model and learns to dynamically control the rate based on the channel condition and image contents. Specifically, a policy network is introduced to exploit the tradeoff space between the rate and signal quality. To train the policy network, the Gumbel-Softmax trick is adopted to make the policy network differentiable and hence the whole JSCC scheme can be trained end-to-end. To the best of our knowledge, this is the first deep JSCC scheme that can automatically adjust its rate using a single network model. Experiments show that our scheme successfully learns a reasonable policy that decreases channel bandwidth utilization for high SNR scenarios or simple image contents. For an arbitrary target rate, our rate-adaptive scheme using a single model achieves similar performance compared to an optimized model specifically trained for that fixed target rate. To reproduce our results, we make the source code publicly available at \url{https://github.com/mingyuyng/Dynamic_JSCC}.
\end{abstract}
\section{Introduction}
\label{sec:intro}
\figstructure
Based on Shannon's separation theorem \cite{shannon1948mathematical}, most modern systems adopt a separate source coding (e.g., JPEG, BPG) and channel coding (e.g., LDPC, Polar, etc.) for wireless image transmission. However, to achieve optimality, the length of the codeword needs to be (infinitely) long. Besides, the optimality also breaks down with non-ergodic source or channl distribution. Thus, using joint source channel coding (JSCC) can enable significant gains as demonstrated in various schemes such as vector quantization and index assignment \cite{vosoughi2014joint, heinen2005transactions, bozantizis2000combined}.  

In recent years, due to the significant performance improvement over traditional analytical methods, deep learning has been applied to various applications such as computer vision \cite{krizhevsky2012imagenet} and wireless communications \cite{ye2017power, yang2019ilps, hsiao2021super}. Because of its ability to extract complex features, deep learning also has been extended to JSCC systems for wireless image transmission \cite{bourtsoulatze2019deep, kurka2019successive, kurka2020deepjscc, yang2021deep, choi2019neural, xu2021wireless} to achieve better performance than separated source and channel coding. In this paper, we focus on the scenario where the communication resource is limited or expensive as in typical Internet-of-Things applications. For such applications, it is desirable that the JSCC scheme dynamically adjusts the rate to save the channel bandwidth while communicating sufficient amount of source information contents. However, most existing deep JSCC methods are only trained with a fixed rate, thus requiring multiple trained models to achieve a multi-rate JSCC scheme. 
Recently, the concept of multi-rate deep JSCC has been investigated in \cite{kurka2019successive, kurka2021bandwidth} where a single model is trained with multiple rates. However, the models in \cite{kurka2019successive, kurka2021bandwidth} are not adaptive to the channel condition as they are only trained with a single SNR. Besides, they do not consider a built-in mechanism or policy that enables automatic rate control given the multi-rate model. 

In this paper, we propose a novel JSCC scheme that adapts its rate conditioned on the channel SNR and image contents with a single network model. To train the JSCC model with various SNRs, we adopt the SNR-adaptive modules introduced in \cite{xu2021wireless} to modulate the intermediate features based on the SNR. Furthermore, motivated by the adaptive computation idea in \cite{figurnov2017spatially, wang2018skipnet, gao2018dynamic, veit2018convolutional}, we introduce a policy network that learns to dynamically decide the number of active features for transmission. We adopt the Gumbel-Softmax trick \cite{jang2016categorical} to make the decision process differentiable. Our experiment results show that our deep learning model can be trained to learn a policy that assigns a various rate for different SNRs and image contents. Specifically, it learns to reduce channel bandwidth utilization when the SNR is high and/or the source image contains less information. At the same time, the proposed scheme maintains the equivalent image quality compared to a specialized model trained for that particular rate. 
\vspace{-0.3cm}
\figadap
\section{Adative JSCC Method}
\label{sec:method}
\vspace{-0.2cm}
\subsection{Overview}
The overall structure of the proposed deep JSCC scheme is shown in Fig. \ref{fig:fig_structure}. First, the source-encoder $E_s$ extracts the image features $X_s$ from a source image $X$. Then $X_s$ is fed to the channel-encoder $E_c$ which generates $(G_s+G_n)$ groups of length-$L$ features. The first $G_s$ groups contain \textit{selective} features that can be either active or inactive as controlled by the policy network, whereas the last $G_n$ groups only have \textit{non-selective} features that are always active. Instead of handcrafting the selection, we introduce a policy network $P$ that outputs a binary mask $W\in\{0, 1\}^{G_s\times 1}$ to select active features for each input $X$. The total number of active groups is denoted as $\tilde{G}=G_n + \sum_{i=1}^{G_s} W_i$. After selection, all active features are passed through the power normalization module to generate complex-valued transmission symbols $X_c\in \mathbb{C}^{\tilde{G}\times L/2}$ with unit average power using the first half of features as the real part and the other half as the imaginary part. $Y_c$ is received as $X_c$ is transmitted over the noisy wireless channel. Then at the receiver, $Y_c$ is fed to the channel-decoder $D_c$ and source-decoder $D_s$ sequentially to reconstruct the source image. 

We quantify the transmission rate in term of wireless channel usage per pixel (CPP). Suppose an input image has $H\times W$ dimension for RGB (3 channel) pixels. Then the CPP is defined as $CPP = \frac{\tilde{G}L}{2HW}$, which is within the range of $[\frac{G_n L}{2HW}, \frac{(G_n+G_s) L}{2HW}]$ depending on the number of active feature groups. The 1/2 factor in CPP is because of complex-valued (quadrature) transmission over the wireless channel. We only consider the AWGN wireless channel such that $Y_c = X_c + N_c$ holds where $N_c$ is the complex Gaussian noise vector. The channel condition/quality is fully captured by the signal-to-noise ratio (SNR), which is assumed to be known at both transmitter and receiver. The SNR value is fed to $E_c$, $D_c$, and $P$ so that the model adapts to the wireless channel condition. 

\vspace{-0.2cm}
\subsection{SNR-adaptive channel encoder and decoder}

The structures of the channel-encoder $E_c$ and channel-decoder $D_c$ networks are shown in Fig. \ref{fig:fig_adaptive}, where additional SNR-adaptive modules are inserted between layers to modulate the intermediate features. The structure of the SNR-adaptive module is inspired by \cite{xu2021wireless} and shown at the bottom of Fig. \ref{fig:fig_adaptive}. The input features are first average pooled across each channel\footnote{We use `channel' to indicate a feature channel of a neural network and `wireless channel' to indicate the wireless transmission channel.} and then concatenated with the SNR value. After that, they are passed through two multi-layer perceptrons (MLP) to generate the factors for channel-wise scaling and addition. In the channel-encoder $E_c$, the input features are first fed to a series of ResNet and SNR-adaptive modules. After that, they are projected to a specific output size through 2D convolution and reshaping. In the channel-decoder $D_c$, since we only receive symbols in $\tilde{G}$ active groups selected by the transmitter, we simply zero-pad the deactivated ones to keep the input size same regardless of the CPP.   
\vspace{-0.2cm}
\subsection{Learning adaptive policy for rate control}

\subsubsection{Policy Network}
The proposed policy network learns to select the number of active feature groups conditioned on the SNR and image content. The whole process can be modeled as sampling a categorical distribution whose sample space is $\Omega = \{0, 1, 2, ..., G_s\}$. The structure of the policy network is shown in Fig. \ref{fig:fig_policy}, where the image feature $X_s$ is first average pooled and concatenated with the SNR. Then, it is passed through a two-layer MLP with a softmax function at the end to generate the probabilities for each option. After that, we sample the decision as an one-hot vector through Gumbel-Softmax (discussed later) and further transform it to a \textit{thermometer-coded} vector as the final adaptive transmission mask $W$. This thermometer encoding of $W$ ensures that we always activate consecutive groups of features from the beginning. Thus, no extra control messages (except the end of transmission signaling) are needed to inform the receiver which groups of features are activated vs. deactivated.   

\figpolicy
\vspace{-0.2cm}
\subsubsection{Gumbel Softmax}

Training the policy network is not trivial because the sampling process is discrete in nature, which makes the network non-differentiable and hard to optimize with back-propagation. One common choice is to use a score function estimate to avoid back-propagation through samples (e.g., REINFORCE\cite{williams1992simple}). However, that approach often has slow convergence issues for many applications and also has a high variance problem \cite{wu2018blockdrop}. As an alternative, we adopt the Gumbel-Softmax scheme \cite{jang2016categorical} to resolve non-differentiability by sampling from the corresponding Gumbel-Softmax distribution. 

Suppose the probability for each category is $p_k$ for $k = 0, ..., G_s$. Then, with the Gumbel-Max trick \cite{jang2016categorical}, the discrete samples from the distribution can be drawn as:
\begin{equation}
     \hat{P} = \argmax_k (\log p_k + g_k), \;\;\; k \in [0, 1, ..., G_s],
    \label{eq:discrete}
\end{equation}
where $g_k=-\log(-\log U_k)$ is a standard Gumbel distribution with $U_k$ sampled from a uniform distribution $U(0,1)$. Since the argmax operation is not differentiable, Gumbel-Softmax distribution is used as a continuous relaxation to argmax. With a goal to represent $\hat{P}$ as a one-hot vector $\bar{P}$, we use a relaxed form $\tilde{P}$ using the softmax function:
\begin{equation}
     \tilde{P}_k = \frac{\exp((\log p_k + g_k)/\tau)}{\sum_{j=0}^{G_s}\exp((\log p_j + g_j)/\tau)}, \;\;\; k = 0, 1, ..., G_s,
    \label{eq:continuous}
\end{equation}
where $\tau$ is a temperature parameter that controls the discreteness of $\tilde{P}$. The distribution converges to a uniform distribution when $\tau$ goes to infinity, whereas $\tau\approx0$ makes $\tilde{P}$ close to a one-hot vector and indistinguishable from the discrete distribution. For the forward pass during network training, we sample the policy from the discrete distribution (\ref{eq:discrete}) whereas the continuous relaxation (\ref{eq:continuous}) is used for the backward pass to approximate the gradient. When the trained model is used for the adaptive-rate JSCC, we transform the one-hot vector output to a thermometer encoded mask vector $W$ satisfying $W_k = \sum_{i=k}^{G_s} \bar{P}_i, \;\;\; k = 1, ..., G_s$. 

\subsection{Loss function}
During training, we minimize the following loss to encourage improving the image reconstruction accuracy as well as minimizing the number of active feature groups to reduce the bandwidth usage (i.e., lower CPP).
\begin{equation}
     \mathbb{E}_{X\sim \mathcal{D}_{train}} \left[ \| X-Y\|_2^2 + \alpha \sum_{i=1}^{G_s}W_i\right]
    \label{eq:loss}
\end{equation}
The first term in (\ref{eq:loss}) is the reconstruction loss and the second term is the channel usage with a weight parameter $\alpha$. 

\section{Experiments}

\label{sec:experiment}
\subsection{Training and implementation details}
We evaluate the proposed method with the CIFAR-10 dataset which consists of 50000 training and 10000 testing images with $32 \times 32$ pixels. The Adam optimizer \cite{kingma2014adam} is adopted to train the model. We first train the networks for 150 epochs with a learning rate of $5\times 10^{-4}$ and another 150 epochs with a learning rate of $5\times 10^{-5}$. Then, we fix the encoder $E_s$ and policy network $P$, and fine-tune the other modules for another 100 epochs. The batch size is 128. The initial temperature $\tau$ is 5 and it gradually decreases with an exponential decay of $-0.015$. During training, we sample the SNR uniformly between 0dB and 20dB. For all experiments, we set $G_n=4$, $G_s=4$, and $L=128$. Thus the proposed method provides 5 possible rates (CPP) in total: $0.25, 0.313, 0.375, 0.438, 0.5$. To serve as the baseline, we also train multiple fixed-rate models that use pre-determined sub-group-level feature masking to obtain a constant target CPP. 

\subsection{Results on CIFAR-10}

\figResRate
\figResPSNR

We first show the trade-off between the average rate and image quality from the proposed method in Fig. \ref{fig:fig_results_rate}. Image quality is evaluated with the peak signal-to-noise ratio (PSNR). The PSNR of baselines with the maximum ($\text{CPP}=0.5$) and minimum ($\text{CPP}=0.25$) fixed rate are plotted in black as a reference. When the SNR is low (0dB), our method tends to select the maximum possible rate. As we gradually increase the SNR, the average rate drops. This trend shows that our method can successfully learn a policy that utilizes less channel resources (lower CPP rate) when the channel condition (SNR) is good and vice versa. As we increase $\alpha$ in the loss function, the policy network pays more attention to the rate than the image quality, and thus the rate decreases faster when SNR increases. 

Next, we compare the performance of the proposed method with 1) the state-of-the-art image codec BPG combined with idealistic error-free transmission based on Shannon capacity, and 2) collection of fixed-rate baseline models each trained for one particular rate. 
The result is shown in Fig. \ref{fig:fig_results_psnr}. 
With a relatively small $\alpha=5\times 10^{-4}$, our method chooses a relatively high CPP (compared to a high $\alpha$ counterpart) given SNR and it outperforms the BPG+Capacity in the low SNR region. Whereas when trained with $\alpha=1.5\times 10^{-3}$, the resulting CPP is relatively small and our method outperforms the BPG+Capacity baseline for all SNRs. 
Compared with baseline models  specifically trained for a fixed rate with a pre-determined activation masking, our method always achieves a similar performance for each comparison point although our method only uses a single trained model adapting to different SNRs.

\figResVar

Finally, we fix the SNR as well as $\alpha$ to observe the variation of the  rate across different image classes. In Fig. \ref{fig:fig_results_variance}, we plot the average rates and PSNR for all 10 classes in CIFAR-10. It can be seen that uneven rate is adopted to different classes because the policy network tends to assign higher CPP to classes with richer information (e.g., Automobile, Truck) while assigning lower CPP to the classes with relatively simple contents (e.g., Ship, Airplane). With such a strategy, our scheme tends to decrease the variation of reconstructed image quality (PSNR) across all classes compared to a fixed-rate baseline which assigns the same CPP to all classes. For this particular SNR and $\alpha$, the standard deviations of PSNR across different classes are 1.017, 0.991, 0.985, 0.947, and  0.923 for five fixed-rate baseline models with CPP = 0.25, 0.313, 0.375, 0.438, and 0.5, respectively. By contrast, our method with adaptive rate control exhibits a significantly lower standard deviation of 0.613. It indicates that when the average CPP is the same, our method produces more equalized image quality across difference classes (with an adaptive rate per class) while the fixed rate scheme generates unbalanced quality images to force the CPP same for all classes.

\section{Conclusion}
\label{sec:conclusion}

In this paper, we present a novel deep JSCC scheme that supports multiple rates with a single model. The policy network automatically assigns the rate conditioned on the channel SNR and image content by dynamically producing a binary mask to activate or deactivate image features. To make the policy network differentiable, the Gumbel-Softmax trick is adopted. Experiments show that our method can learn a reasonable policy that distributes less bandwidth when the SNR is high and the image contains less information. With the advantage of adaptive rate control, our method only experiences negligible performance degradation compared with multiple single-rate models at each  operating condition.     

\newpage
\bibliographystyle{IEEEbib}
\bibliography{refs}

\end{document}